# Electronic properties of novel 6K superconductor LiFeP

I. R. Shein* and A. L. Ivanovskii

*Institute of Solid State Chemistry, Ural Branch, Russian Academy of Sciences, Pervomaiskaya St., 91, Ekaterinburg, 620990 Russia*

**A B S T R A C T**

Very recently, the new 6K superconductor (SC) LiFeP, the first arsenic-free analogue of the family of the so-called "111" FeAs SCs, was discovered. Here, based on first-principle FLAPW-GGA calculations, the band structure, density of states, Fermi surface topology, electron density distribution and effective atomic charges for the new SC LiFeP are investigated and discussed in comparison with isostructural and isoelectronic LiFeAs.

## 1. Introduction

Following the discovery of superconductivity at $T_C \sim 26K$ in LaFeAsO(F) [1], spirited search has led to a broad family of FeAs-based superconductors (SCs), see reviews [2-4]. Among them, five main groups of related FeAs SCs known as «1111», «122», «111», «32225» and «42226» materials have been found to date. The parent phases for these groups of FeAs SCs are $Ln$FeAsO and $B$FeAsF for "1111", $B$Fe$_2$As$_2$ for "122", $A$FeAs for "111", Sr$_3$Sc$_2$Fe$_2$As$_2$O$_5$ for «32225» and Sr$_4$M$_2$Fe$_2$As$_2$O$_6$ for «42226», where $Ln$, $A$, $B$ and $M$ are rare earth, alkaline, alkaline earth and transition metals, respectively.

All the above FeAs SCs adopt a quasi-two-dimensional crystal structure, where [$Fe_2As_2$] blocks are separated either by atomic $A$, $B$ sheets (for three-component "111" and "122" phases) or by [$Ln$O], [$B$F] blocks (for four-component "1111" phases) or by more complex perovskite-like blocks (for five-component «32225» and «42226» phases). For all of the FeAs SCs, the electronic bands in the window around the Fermi level are formed mainly by the states of the [$Fe_2As_2$] blocks and play the main role in superconductivity, whereas the mentioned atomic sheets or blocks serve as "charge reservoirs", see [2-4].

Besides superconductivity, these materials possess various interesting physical properties, such as ferromagnetism, antiferromagnetism, spin density waves (SDW) *etc*. However, although the understanding of the nature of coexistence of these properties is of crucial importance, it is still under debate.

In this context, the effects of chemical substitutions (especially in conducting blocks [$Fe_2As_2$]) on superconductivity, magnetic and other physical properties of the above materials are of great interest. In particular, partial replacement of As (in [$Fe_2As_2$] blocks) by phosphorus in "122" phases leads to suppression of SDW instability in undoped crystals, and the highest superconducting transitions are observed at $T_C$ ~ 30 K in $BaFe_2As_{1.36}P_{0.64}$ [5], $T_C$ ~27 K in $SrFe_2As_{1.3}P_{0.7}$ and $T_C$ ~13 K in $CaFe_2As_{1.7}P_{0.3}$ [6]. Moreover, recently a rich set of isostructural analogues of FeAs SCs, where ***arsenic is completely replaced by phosphorus,*** was synthesized. For example, among them there are $Ln$FePO, where $Ln$ = La, Pr and Nd (belong to the above mentioned "1111" family) [7], $SrFe_2P_2$ [8], $BaFe_2P_2$ [5] (belong to the "122" family), and $Sr_4Sc_2Fe_2P_2O_6$ (belongs to the "42226" family) [9]. Empirically, all these phosphorus-containing materials are non-SCs or show low - $T_C$ superconductivity (with $T_C$ < 5 K, see [2-9]), except $Sr_4Sc_2Fe_2P_2O_6$ with $T_C$ ~ 17 K [9] – the highest transition temperature among those reported for phosphorus-containing systems.

Very recently, the new 6K SC LiFeP was discovered [10], which is the first arsenic-free analogue of the family of the so-called "111" FeAs SCs.

In this Communication, by means of first-principle FLAPW-GGA calculations, we studied the electronic properties for the newly discovered LiFeP with the purpose to evaluate the peculiarities of its band structure, density of states, Fermi surface topology, electron density distribution and effective atomic charges in comparison with isostructural and isoelectronic 18K SC LiFeAs.

## 2. Structural models and computational aspects

LiFeP crystallizes in a tetragonal unit cell, space group P4/*nmm*. This quasi-two-dimensional structure is built up of [Fe$_2$P$_2$] blocks alternating along the *c* axis with nominal double layers of Li atoms. The same structure is adopted by the known LiFeAs "111" phase, see [11-15]. As the results of atomic positions determination for LiFeAs remain debatable [11-15], in our comparative study of LiFeP and LiFeAs, we used a uniform structural model according to Ref. [11], see Table 1. For this purpose, we used the experimental lattice parameters *a*=*b*=3.692 Å; *c*=6.031 Å for LiFeP [10] and *a*=*b*=3.791 Å; *c*=6.364 Å for LiFeAs [11] and optimized internal coordinates $z_{Li}$ and $z_{P,As}$. These self-consistent calculations were considered to be converged when the difference in the total energy of the crystal did not exceed 0.1 mRy and the difference in the total electronic charge did not exceed 0.001 *e* as calculated at consecutive steps. The obtained data are $z_{Li}$ =0.8395 and $z_P$ = 0.2024 for LiFeP and $z_{Li}$ = 0.8289 and $z_{As}$= 0.2897 for LiFeAs (for this compound the experimental data are $z_{Li}$ = 0.8459 and $z_{As}$= 0.2635 [11]).

Our calculations were carried out by means of the full-potential method with mixed basis APW+lo (LAPW) implemented in the WIEN2k suite of programs [16]. The generalized gradient correction (GGA) to exchange-correlation potential in the PBE form [17] was used. The plane-wave expansion was taken to $R_{MT} \times K_{MAX}$ equal to 7, and the $k$ sampling with 11×11×7 $k$-points in the Brillouin zone was used.

## 3. Results and discussion

Figures 1 and 2 show the band structures, total and atomic-resolved $l$-projected DOSs for LiFeP and LiFeAs; the calculated values of bandwidths are presented in Table 1.

For LiFeP, the two lowest bands lying around -11 eV below the Fermi level ($E_F$) arise mainly from P 3$s$ states (peak A Fig. 2) and are separated by a gap (~ 4.3 eV) from the near-Fermi valence bands, which are located in the energy range from -5.9 eV to $E_F$ and are formed predominantly by Fe 3$d$ and P 3$p$ states. The corresponding total DOS include two main subbands B and C, Fig. 2. The subband B contains strongly hybridized Fe 3$d$ - P 3$p$ states, which are responsible for the covalent Fe-P bonding, see also below. The intense peak C in the DOS is due to the Fe 3$d$-like bands with low $E(k)$ dispersion, which are located around -1 eV; these states participate in metallic-like Fe-Fe bonds. Finally, the bottom of the conduction band (subband D) is also made up basically of Fe 3$d$ states with an admixture of anti-bonding P 3$p$ states. Thus, the near-Fermi region is formed mainly by the states of [Fe$_2$P$_2$] blocks. Besides, it is noteworthy that the contributions from the valence states of Li to the occupied subbands are quite negligible, *i.e.* in LiFeP (as well as in LiFeAs) lithium atoms are in ionized forms close to cations Li$^{1+}$.

In general, the band structure and DOSs distributions for LiFeAs (Figs. 1, 2) match those for LiFeP, and these results are also in good agreement with the data of earlier calculations for this crystal, see [18-22].

The Fermi surfaces (FSs) for LiFeP and LiFeAs are depicted in Fig. 3. They are very similar, adopt a two-dimensional character typical of FeAs SCs [2-4], and consist of a system of sheets parallel to the $k_z$ direction, where concentric hole-like cylinders are at the zone center ($\Gamma$) and electron-like sheets are centered along the *M-A* direction in the corners of the Brillouin zone.

For further description of the differences in the electronic structures for LiFeP *versus* LiFeAs it should be taken into account that these compounds are isostructural and isoelectronic, but when going from LaFeAs to LaFeP, the lattice parameters decrease.

In case of the mentioned iron-pnictogen covalent bonds, the bonding-antibonding splitting should be very sensitive to the distance between these atoms. Thus, the reduction in distances Fe-P *versus* Fe-As leads to an increase in bonding-antibonding splitting for LiFeP and to moving of Fe 3*d* states away from the Fermi level. As a result, the total DOS at the Fermi level N($E_F$) (with the dominant contribution from Fe 3*d* states) for LiFeP becomes lower than for LiFeAs, see Table 3, where we also present our estimations for the Sommerfeld constants ($\gamma$) and the Pauli paramagnetic susceptibility ($\chi$) for these phases under the assumption of the free electron model as $\gamma = (\pi^2/3)N(E_F)k_B^2$ and $\chi = \mu_B^2 N(E_F)$. Note that the available experimental Sommerfeld constants as obtained from specific-heat measurements [10, 23] also demonstrate a decrease in the sequence $\gamma$(LiFeAs) > $\gamma$(LiFeP), in agreement with our estimations.

One of the most intriguing peculiarities of FeAs SCs is the presence of typically metallic collective excitations, such as itinerant magnetization waves in the non-superconducting undoped parent "1111" or "122" phases, for which at least three different competing types of magnetic fluctuations have been predicted, see [2-4]. Within our band structure calculations, comparative magnetic instability of LiFeP *versus* LiFeAs may be examined by the simple Stoner criterion, according to which magnetism may occur if N($E_F$)$I$ > 1. Here N($E_F$) is the density of states at the Fermi level on an atom per spin basis. Taking the typical value of $I$ = 0.9 [24], our estimations show that the parameter N($E_F$)$I$ changes from 0.75 (for LiFeP) to 0.89 for LiFeAs, which becomes most unstable to magnetism among these "111" materials.

Let us discuss the inter-atomic bonding in LiFeP in comparison with LiFeAs in more detail. Our results show that like for the related «1111», «122», «111», «32225» and «42226» materials [25-29], the general bonding picture in LiFeP and LiFeAs is very similar and can be classified as a high-anisotropic mixture of ionic, covalent and metallic contributions.

Indeed, inside [Fe$_2$P$_2$(As$_2$)] blocks the metallic-like Fe-Fe bonds appear owing to delocalized near-Fermi Fe 3$d$ states, see Fig. 2.

To describe the ionic bonding, we start with a simple ionic picture, which considers the usual oxidation numbers of atoms: Li$^{1+}$, Fe$^{2+}$, and (P,As)$^{3-}$. Thus, the charge states are (1+) for single atomic Li sheets and (2-) for blocks [Fe$_2$P$_2$(As$_2$)], *i.e.* the charge transfer occurs from Li$^{1+}$ sheets to [Fe$_2$P$_2$(As$_2$)]$^{2-}$ blocks. Besides, inside [Fe$_2$P$_2$(As$_2$)] blocks, the ionic bonding takes place between Fe-P(As) atoms.

For numerical estimation of the amount of electrons redistributed between the adjacent Li sheets and [Fe$_2$P$_2$(As$_2$)] blocks and between Fe and P(As) atoms inside these blocks, we carried out a Bader [30] analysis. The effective atomic charges defined as $\Delta Q = Q^B - Q^i$ (where $Q^B$ and $Q^i$ are the so-called Bader charges and the charges as obtained from the purely ionic model, respectively) are presented in Table 4. These results clearly demonstrate that the inter-atomic as well as inter-layers charge transfer is smaller than predicted in the idealized ionic model. Namely, the transfer $\Delta Q(\text{Li} \rightarrow [\text{Fe}_2\text{P}_2(\text{As}_2)])$ is about 0.86 $e$. Note also that the charge transfer between atoms inside [Fe$_2$P$_2$(As$_2$)]) blocks (for example, for LiFeP: from Fe to P at about 0.42 $e$) is much smaller than between the adjacent Li sheets and [Fe$_2$P$_2$(As$_2$)] blocks. This implies that individual ionic bonds between various atoms and layers are highly anisotropic. In addition, the transfer from Fe to pnictogen is smaller for LiFeAs than for the phosphide, Table 4. This fact can be explained by higher electronegativity of phosphorus and by contraction of Fe-P distances in comparison with Fe-As distances. Note that the same tendency was established for FeAs "122" phases in comparison with their P-containing counterparts [25].

The character of covalent bonding in LiFeP and LiFeAs phases may be well understood from site-projected DOS calculations. As is shown in Fig. 2, Fe-P(As) states are strongly hybridized. These covalent bonds are clearly visible also in Fig. 4, where the charge density map for LiFeP is depicted. Besides, there are no directed bonds between Li sheets and [Fe$_2$P$_2$] blocks, *i.e.* in contrast to "122" phases [25, 29], the inter-layer bonding in LiFeP (as well as in LiFeAs) is purely ionic. Thus, summarizing the above results, the examined materials may be described as *ionic metals*.

**Conclusions**

In conclusion, we used the first-principle FLAPW-GGA approach to investigate the band structure, density of states, Fermi surface topology, electron density distribution and effective atomic charges for the recently synthesized 6K superconductor LiFeP - the first arsenic-free analogue of the family of the so-called "111" FeAs SCs.

Our results show that similar to other "111" FeAs phases, for LiFeP the density of states in the vicinity of the Fermi energy is found to be dominated by contributions from the Fe 3$d$ states. The main differences in the electronic structures for isostructural and isoelectronic phases LiFeP and LiFeAs are due to reduction in inter-atomic distances for the iron-phosphide phase, resulting in an increase in the bonding-antibonding splitting. The picture of inter-atomic bonding for LiFeP may be described in the following way: (i) inside [Fe$_2$P$_2$] blocks, mixed covalent-ionic bonds Fe-P take place (owing to hybridization of Fe 3$d$ – P 3$p$ states and Fe → P charge transfer); (ii) inside [Fe$_2$P$_2$] blocks, metallic-like Fe-Fe bonds appear owing to delocalized near-Fermi Fe 3$d$ states; (iii) between the adjacent [Fe$_2$P$_2$] blocks and Li sheets, ionic bonds emerge owing to Li → [Fe$_2$P$_2$] charge transfer.

Finally, in view of the results obtained, it is reasonable to hypothesize that the structural changes, which accompany the replacement of arsenic in LiFeAs by chemically similar but slightly smaller phosphorus, can be considered as one of the main factors responsible for the experimentally observed lowering of $T_C$ as going from LiFeAs ($T_C$ ~18K) to LiFeP ($T_C$ ~6K). Indeed, in comparison with LiFeAs, its phosphorus-containing analogue LiFeP can be viewed as a "compressed" "111" phase. In turn, examination of the effect of hydrostatic pressure on superconductivity in LiFeAs shows that for this material $T_C$ decreases linearly with pressure at a rate of ~ 1.5K/GPa [31,32]. Thus, it may be supposed that $T_C$ of LiFeP can be enhanced by a negative pressure or by tensile strain. For example, a negative pressure for LiFeP can be achieved by replacement of the smallest Li ions by ions with large radii in the alkali metal series (Na, K etc), whereas tensile strain can be achieved by deposition of LiFeP thin films on a substrate with an appropriate lattice.

In our opinion, further experimental efforts to check up the above assumption are of interest: unlike the previous doping strategy of changing the concentration of carriers for iron-pnictogen SCs, the proposed approach does not alter the number of valence electrons.

**Acknowledgments**

Financial support from the RFBR (Grant 09-03-00946-a) is gratefully acknowledged.

**Table 1**.
Atomic positions for LiFeP and LiFeAs. Space group 129, cell choice 2.

| atom | site | $x$ | $y$ | $z$ |
|---|---|---|---|---|
| Li | 2c | ¼ | ¼ | $z_{Li}$ |
| Fe | 2b | ¾ | ¼ | ½ |
| P (As) | 2c | ¼ | ¼ | $z_{As,P}$ |

**Table 2**.
Calculated values of bandwidths (in eV) for for LiFeP and LiFeAs.

| system | LiFeP | LiFeAs |
|---|---|---|
| Common bandwidth | 12.4 | 12.5 |
| Valence band (Fe 3$d$+ P(As) n$p$) | 5.8 | 5.5 |
| Bang gap | 4.3 | 5.0 |
| Quasi-core pnictogen $s$ band | 2.3 | 2.0 |

**Table 3**.
Total (in states/eV per cell) and partial densities of states at the Fermi level (in states/eV per atom), electronic heat capacity γ (in mJ·K$^{-2}$·mol$^{-1}$) and molar Pauli paramagnetic susceptibility χ (in 10$^{-4}$ emu/mol) for LiFeP and LiFeAs.

| system | P(As) n$p$ | Fe 3$d$ | total | γ | χ |
|---|---|---|---|---|---|
| LiFeP | 0.037 | 1.356 | 3.349 | 7.894 | 1.078 |
| LiFeAs | 0.028 | 1.629 | 3.823 (3.86 [22]; ~4.0 [24]; 4.2 [36]) | 9.011 | 1.231 |

\* other available theoretical data are given in parentheses.

**Table 4.**
Effective atomic charges (ΔQ, in $e$) for LiFeP and LiFeAs as obtained from Bader analysis.

| system | Li | Fe | P(As) |
|---|---|---|---|
| LiFeP | +0.858 | +0.420 | -1.278 |
| LiFeAs | +0.857 | +0.174 | -1.031 |

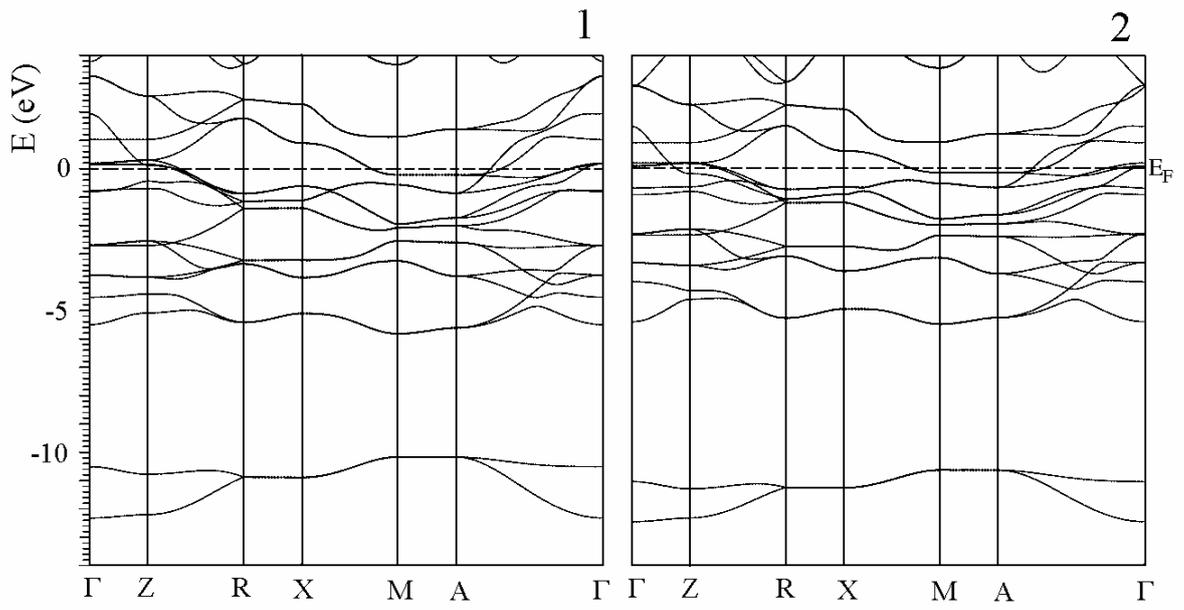

**Figure 1.** Electronic band structures of LiFeP (1) and LiFeAs (2).

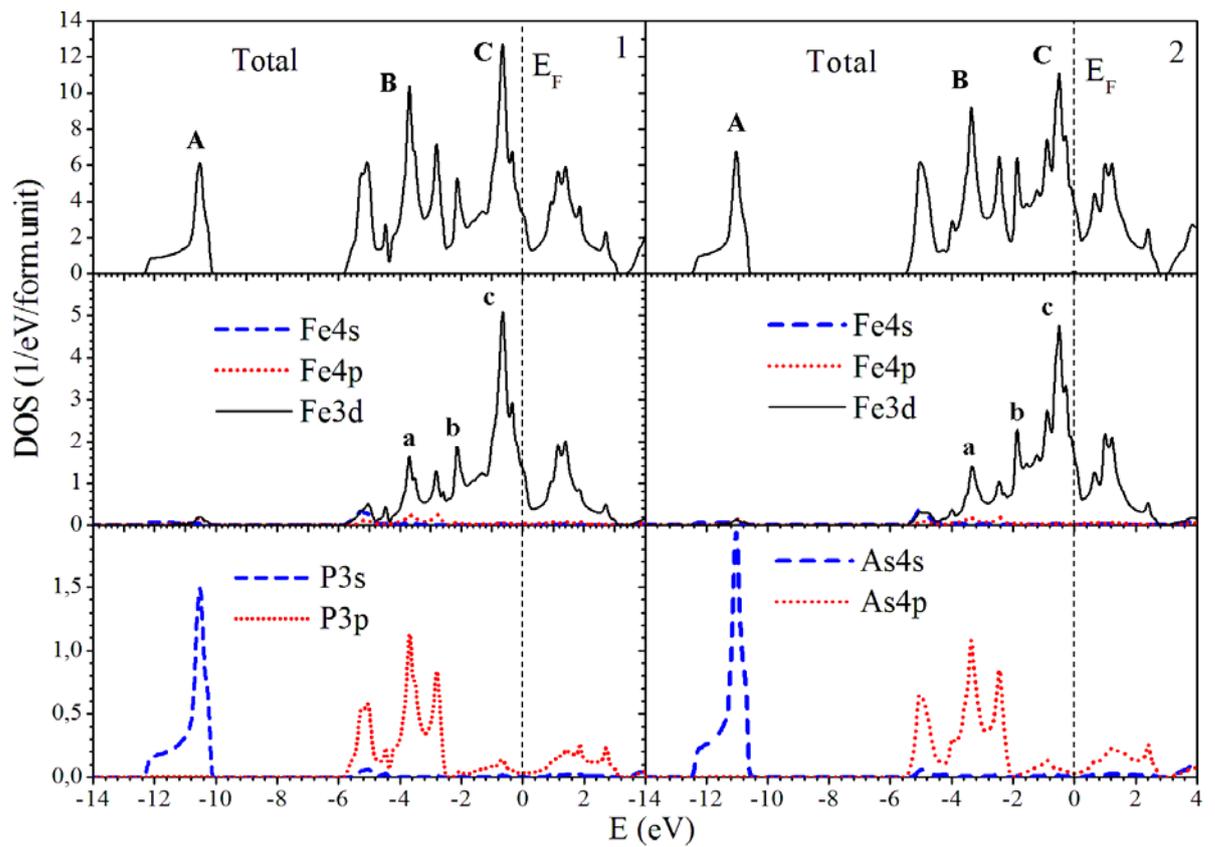

**Figure 2.** Total and partial densities of states of LiFeP (1) and LiFeAs (2).

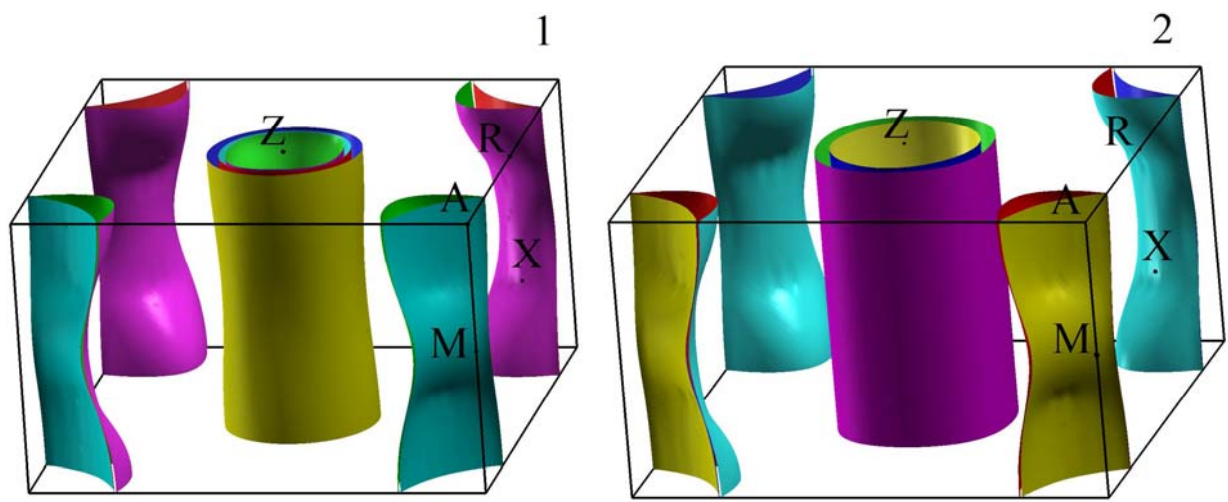

**Figure 3**. The Fermi surfaces of LiFeP (1) and LiFeAs (2).

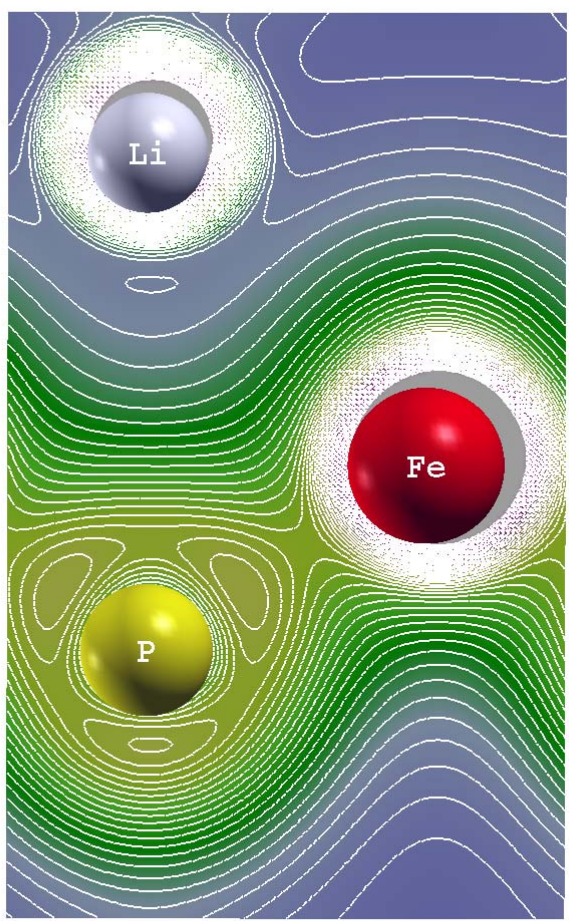

**Figure. 4**. Valence charge density map in [100] plane for LiFeP.